\RequirePackage{fix-cm}
\documentclass[twocolumn,epjc3]{svjour3}

\smartqed

\RequirePackage{graphicx}

\RequirePackage{amsmath,amsfonts,amssymb}
\RequirePackage{dsfont}
\RequirePackage{graphicx}
\RequirePackage{subfigure,float}

\journalname{Eur. Phys. J. C}

\begin{document}
\title{A universe born in a metastable false vacuum state\\ needs not die}

\author{K. Urbanowski }

\institute{University of Zielona G\'{o}ra, Institute of Physics, \\
ul. Prof. Z. Szafrana 4a,  65--516 Zielona G\'{o}ra, Poland. \label{addr1}
}

\thankstext{e1}{e-mail:  K.Urbanowski@if.uz.zgora.pl,\\ k.a.urbanowski@gmail.com}

\date{Received: date / Accepted: date}

\maketitle

\begin{abstract}

We try to find  conditions, the fulfillment of which allows a universe born in a metastable false vacuum state to survive and not to collapse.
The conditions found are in the form of inequalities linking the depending on time $t$ instantaneous decay rate ${\it\Gamma}(t)$ of the false vacuum state and the Hubble parameter $H(t)$. Properties of the decay rate of a quantum metastable states are discussed and then the possible solutions of the conditions found are analyzed and discussed.
Within the model considered it is shown that  a universe born in the metastable vacuum state  has a very high chance of surviving until very late times
if the lifetime, $\tau_{0}^{F}$, of the metastable false vacuum state is much shorter, than the duration of  the inflation process.
Our analysis shows that the instability of the electroweak vacuum does not have to result in the tragic fate of our Universe leading to its death.
\end{abstract}


\section{Introduction}

Various aspects of the problem of whether the Universe can live forever or whether sooner or later will have to decay (i. e. to die) was considered in mathematical cosmology based on solutions to Einstein's equations practically from the very beginning (see, eg. \cite{Ed,lem1} and also a great review works \cite{spiros1,spiros2}).
This discussion gained a new face and intensified after the publication of a seminal series of papers by  Coleman et al. \cite{Coleman:1977py,Callan:1977pt,Col2},
where the instability of a physical system, which is not at an absolute energy minimum, and which is separated from the absolute minimum by an effective potential barrier
was discussed. They
showed that if the early Universe is too cold to activate the classical  energy  transition
(i. e. the transition of  the form of classical thermally activated  jumps over the barrier)
to the minimum energy state
then a quantum decay, from the false vacuum to the true vacuum, is still possible through a barrier penetration via the
quantum tunneling.
In other words they showed that quantum decay processes can play an important role in the early Universe and
therefore we will assume that a creation of the Universe is a quantum process.
The results presented in \cite{Coleman:1977py,Callan:1977pt,Col2} mean
that all early universes, in which the the state describing their lowest energy
was the  quantum metastable vacuum state,
may decay and die sooner or later.
 (We assume that other parallel universes may have been created with our universe).
 Such a possibility of a tragic fate of the Universe has been  discussed in many cosmological models (see e. g. \cite{Af,Kennedy:1980cj,Ta,La,Ba,Al,Stojk,winitzki,Krauss:2007rx,Krauss:2008pt,Sh,Branchina1,Branchina2,Landim:2016isc} and also \cite{Stojk1,maxim,max2,max3,max4,dym} and others).
Some authors analyzing a fate of the Universe born in a metastable false vacuum state
focus on the search for models in which the lifetime, $\tau^{F}_{0}$, of the false vacuum is much longer than the age of our Universe.
Unfortunately, such models may not solve the problem of the possible decay of the Universe
and, as it will be shown in this paper,
such models may generate other problems that are difficult to solve.
Simply, from
the point of view of the quantum theory of unstable states the
long lifetime
of a unstable particle
does not mean that its decay is impossible
(and that
there are no decays)  before that time passes:
The decay probability is small for times shorter than $\tau^{F}_{0}$ but it still  is nonzero.
Therefore
the very long lifetime, $\tau_{0}^{F} $, of the metastable false vacuum state needs not guarantee that the universe with such a vacuum will survive at least until $ \tau^{F}_{0}$ and
that such a universe considered as a quantum unstable particle
will certainly not collapse earlier (or even much earlier) than the time comparable to or longer than the lifetime $ t\geq \tau^{F}_{0}$ will pass. This can happen
unless another mechanism or interactions block that decay.
On the other hand, there are also cosmological models under study in which the lifetime of a false vacuum is very short, and even significantly shorter
than the duration of the inflationary phase.
In \cite{Krauss:2007rx,Krauss:2008pt}, a hypothesis was formulated that regions of the space with such a fale vacuum could survive well until very late times. The same conclusion concerns universes with such a vacuum.

The study of cosmological models with the meta\-sta\-ble false vacua gained additional, even more important significance after the discovery of the Higgs boson and the determination of its mass $m_{H}$ \cite{Ade,Chat}. Estimations of the Higgs and top quark masses \cite{pdg-2020} $m_{H} \simeq 125.10 \pm 0.14$ [GeV] and $m_{t} \simeq 172.76 \pm 0.30$ [GeV]  based on the experimental results  place the Standard Model in the region of the  metastable vacuum (see e. g. \cite{Deg,But,Isidori,bamba,graef,Spencer,Kob,Esp,Elias,Wei,Ema}). A consequence of the  metastability of the Higgs vacuum  is that it should
induce the decay of the electroweak vacuum in the early Universe with catastrophic consequences for the Universe (for a discussion see e.g. \cite{Mar}).
 Considering all of these, the question arises:
 What does prevent of a decay of our Universe?
 Probably for this reason  various mechanisms inducing the electroweak decay process of false vacuum, or slowing it down
 or even stopping it,
 have been discussed in many papers
 (see e. g.  \cite{Mar,Stojk2} and references therein).
Here we try to clarify this issue
from the point of view of the quantum theory of unstable states based on the fundamental assumptions of the quantum theory acting at the interface with Einstein's classical theory of gravity.
Such an
approach frees us from the study of various mechanisms of an instability (or a stability) in the Higgs potential in the Standard Model
and
allows us to draw general conclusions that depend only on whether the false vacuum is a stable or a unstable quantum state.

Properties of a universe born in metastable false vacuum state having very short lifetime have been analyzed in \cite{ku2022}.
In the analysis performed  therein  the model of the dark anergy close to that considered by Landim and Abdalla \cite{Landim:2016isc},
in which the observed vacuum energy is the value of the scalar potential at the false vacuum  was assumed. (Similar idea was used and discussed in many papers --- see eg.
\cite{Stojk,Lim,Sola1,Sola2}). In  other words, it was assumed that
the current stage of accelerated expansion of the universe  is described by a canonical scalar field $\Phi$  such that its potential, $V(\Phi)$, has a local and true minimums. So, the field at the false vacuum represents the dark energy. In such a situation,
the quantum state of the system in the local minimum is described by a state vector corresponding to the metastable false vacuum state whereas the quantum state of the system in the true minimum corresponds to the state of the lowest energy of the system and it is a true vacuum.
This means that the density of the energy of the system in the false vacuum state, $\rho_{vac}^{\,\text{F}}$, is identified with the density of the dark energy,
$\rho_{vac}^{\,\text{F}} \equiv \rho_{de}$, (or, equivalently, as the cosmological term $\Lambda$) in the Einstein equations \cite{Weinberg-cos,Cheng}.

This paper complements the considerations and analysis presented in \cite{ku2022} and  the same assumptions as in \cite{ku2022}  were adopted herein wherever necessary.
Here we propose
conditions those must be met by a universe born in a metastable false vacuum state so that it does not collapse and die.

The paper is organized as follows: In Section 2 an effective volume of an expanding  universe born in a metastable fale vacuum state is discussed and a possible criterion for the survival of such a universe  is proposed. Properties of the decay rate of a quantum unstable state and energy of the system in such a state are analyzed in Sec. 3.
In Sec. 4 we describe a relation  between  some cosmological parameters and quantities discussed in Sec. 3. In Sec.5 an attempt is made to find solutions to the "survival" criterion formulated in Sec. 2. Section 6 contains final remarks.

\section{An effective volume of an expanding  Universe born in a metastable fale vacuum state }
Krauss's and Dent's  statement  in \cite{Krauss:2007rx} that {\em "It may occur
even though some regions of false vacua by assumption
should  decay exponentially, gravitational effects
force space in a region that has not decayed yet
to grow exponentially fast"} may be generalized and expressed in the general case in a mathematical form in terms of the
rate, ${\it\Gamma}(t)$, of the decay process,
and of the rate of the volume expansion.
So, let $V_{eff}(t)$ denote an effective volume of a universe born in the metastable false vacuum state.  According to Krauss's and Dent's hypothesis
this volume is the effect of two independent opposing  processes taking place simultaneously: One of them is the quantum process of a decaying false vacuum characterized by the survival probability ${\cal P}(t)$, the other process is the process of growing the volume, $V(t)$, of the universe depending on the properties of the parameter $\Lambda (t)$ in Einstein's equations, which
 with the Robertson---Walker metric
in the standard form of Friedmann equations \cite{Cheng,Sahni}
look as follows:
The first one,
\begin{equation}
\frac{{\dot{a}}^{2}(t)}{a^{2}(t)}  + \frac{kc^{2}}{R_{0}^{2}\,a^{2}(t)} = \frac{8\pi G_{N}}{3}\,\rho +\frac{\Lambda\,c^{2}}{3}, \label{Fr1}
\end{equation}
and the second one,
\begin{equation}
\frac{{\ddot{a}}(t)}{a(t)} =\,-\,\frac{4\pi G_{N}}{3}\,\left(\frac{3p}{c^{2}} +  \rho \right) + \frac{\Lambda\,c^{2}}{3}. \label{Fr2}
\end{equation}
where "dot" denotes the derivative with respect to time $t$, ${\dot{a}}(t) = \frac{d a(t)}{dt}$,
$\rho$ and $p$ are mass density  and pressure respectively, $k$ denote the curvature signature, and $a(t) = R(t)/R_{0}$ is the scale factor, $R(t)$ is the proper distance at
epoch $t$, $R_{0}= R(t_{0})$ is the distance at the reference time $t_{0}$,
(it can be also interpreted as the radius of the Universe now)  and here $t_{0}$ denotes  the present epoch. The pressure $p$ and the density $\rho$ are
are related to each other through the equation of state, $p = w\rho \,c^{2}$, where $w$ is constant \cite{Cheng}. There is $w=0$ for a dust (for a matter dominated era), $w=1/3$ for a radiation and $w = -1$ for a
vacuum energy.

The simplest possibility is to define the effective volume, $V_{eff}(t)$, as follows
\begin{equation}
V_{eff}(t) = {\cal P}(t)\,V(t),  \label{v-ef}
\end{equation}
where  $V(t) = \frac{4\pi}{3}[R(t)]^{3} = \frac{4\pi}{3}[R_{0}\,a(t)]^{3}$.
Then the rate, $\nu_{V}(t)$,  of the expansion of the volume,  $V_{eff}(t)$, equals
\begin{equation}
\nu_{V}(t) \stackrel{\rm def}{=} \frac{\dot{V}_{eff}(t)}{V_{eff}(t)} = \frac{d\,[{\cal P}(t)\,V(t)]}{dt}, \label{d-r}
\end{equation}
 which gives
\begin{equation}
\nu_{V}(t) = \frac{\dot{V}_{eff}(t)}{V_{eff}(t)} \equiv 3\frac{\dot{a}(t)}{a(t)}\, - \frac{{\it\Gamma}(t)}{\hbar}, \label{d-r1}
\end{equation}
where,
\begin{equation}
\frac{{\it\Gamma}(t)}{\hbar} \,= \,-  \, \frac{1}{{\cal P}(t)}\,\frac{d {\cal P}(t)}{dt}, \label{G1}
\end{equation}
 is the  rate of the decay process. Now if $\nu_{V}(t) \geq 0$ then the
 process  of the expansion
 overcomes the process of the collapse of  the universe
 and as  a result the universe does not have to decay.
 On the other hand if  $\nu_{V}(t)  < 0$  then the fate of such a universe is doomed, and sooner or later it must collapse  and die.

\section{Quantum metastable states: decay rate \\ and related quantities}

We assume that the decay of the metastable false vacuum is the quantum decay process. In general, in such a situation the survival probability equals,
\begin{equation}
{\cal P}(t) = |{\cal A}(t)|^{2}, \label{P(t)}
\end{equation}
where
\begin{equation}
{\cal A}(t) = \langle \phi| \phi(t)\rangle \equiv \langle\phi|e^{\textstyle{-\frac{i}{\hbar}\mathfrak{H} t}}|\phi\rangle,
\label{amp}
\end{equation}
is the survival amplitude, $|\phi\rangle$ is the initial metastable state
and $| \phi (t)\rangle $ is the solution of the
time depended
Schr\"{o}dinger equation with initial conditions $| \phi (t = t_{0}^{init} \equiv 0) \rangle \stackrel{\rm def}{=}
| \phi\rangle$.
Here $\mathfrak{H}$
  denotes the complete (full), self-adjoint Hamiltonian of the system acting in the Hilbert space ${\cal H}$ of states of
  this system, $|\phi\rangle, |\phi (t) \rangle \in {\cal H}$, $\langle \phi|\phi\rangle = \langle \phi(t)|\phi (t) \rangle = 1$
and therefore ${\cal A}(0) = 1$.

Note that the important formula, equivalent to (\ref{G1})), for the decay rate, strictly speaking for the instantaneous decay rate ${\it\Gamma}(t)$, can be found using the equation (\ref{P(t)}). Namely differentiating the product $|{\cal A}(t)|^{2} = {\cal A}(t) [{\cal A}(t)]^{\ast}$ with respect to time $t$ one finds that (see e. g. \cite{ku1993,Urbanowski:1994epq} or \cite{ku2022,jcap-2020} and references therein),
\begin{equation}
{\it\Gamma}(t) = - 2 \Re\,[h(t)] , \label{G2}
\end{equation}
where $\Im\,(z)$ denotes the  imaginary parts of a complex number $z$, and
\begin{equation}
h(t) = i\hbar \frac{1}{{\cal A}(t)}\,\frac{\partial {\cal A}(t)}{\partial t} \equiv \frac{\langle \phi| \mathfrak{H}|\phi(t)\rangle}{{\cal A}(t)}, \label{h(t)}
\end{equation}
is the effective Hamiltonian.
The effective Hamiltonian $h(t)$ governs the time evolution in the subspace of unstable states ${\cal H}_{\parallel}= \mathbb{P} {\cal H}$, where $\mathbb{P} = |\phi\rangle \langle
\phi|$ in the case  considered (see \cite{Urbanowski:1994epq} and also \cite{Urbanowski:2006mw,Urbanowski:2008kra,Urbanowski:2009lpe} and references therein):
\begin{equation}
i\hbar \frac{\partial {\cal A}(t)}{\partial t} = h(t) {\cal A}(t). \label{h-S}
\end{equation}
The subspace ${\cal H} \ominus {\cal
H}_{\parallel} = {\cal H}_{\perp} \equiv \mathbb{Q} {\cal H}$ is the subspace of decay products. Here $\mathbb{Q} = \mathbb{I} - \mathbb{P}$. One meets the effective Hamiltonian
$h(t)$ when one starts from the Schr\"{o}dinger equation for the total state space ${\cal H}$ and looks for the rigorous evolution equation for a distinguished subspace of states
${\cal H}_{||} \subset {\cal H}$ \cite{Urbanowski:1994epq,Giraldi:2015ldu,Giraldi:2016zom}. In general $h(t)$ is a complex function of time and
in the case of ${\cal H}_{\parallel}$ of two or more dimensions
the effective Hamiltonian governing the time evolution in such a subspace is a non-hermitian matrix $H_{\parallel}$ or non-hermitian operator. There is
\begin{equation}
h(t) = E(t) - \frac{i}{2} {{\it\Gamma}}(t), \label{h-m+g}
\end{equation}
where
$ E(t) = \Re\,[h(t)]$, ${\it\Gamma}(t) = -2\,\Im\,[h(t)]$,
are the instantaneous energy (mass), $E(t)$, and the instantaneous decay rate, ${\it\Gamma}(t)$. (Here $\Re\,(z)$ denotes the part of $z$).
In the case of the metastable false vacuum state the instantaneous energy $E(t)$ allows
one to analyze early as well as late properties of the energy density (and thus of the cosmological "constant" $\Lambda(t)$) in the false vacuum state (see \cite{ku2022} and references therein).

In some cases it is convenient to consider the survival amplitude ${\cal A}(t)$ as  a solution of the Schr\"{o}dinger--like equation (\ref{h-S}) with the initial condition ${\cal A}(0) = 1$. Then
\begin{equation}
{\cal A}(t) = e^{\textstyle{-\frac{i}{\hbar}\int_{0}^{t}h(\varsigma) d\varsigma}} \equiv
e^{\textstyle{- \frac{i}{\hbar}\left[\overline{E(t) }- \frac{i}{2} \overline{{\it\Gamma}(t)}\right]t}}, \label{A1}
\end{equation}
where   $\overline{E(t)} = \frac{1}{t}\int_{0}^{t}E(\varsigma)d\varsigma$ is the effective energy of the
system in the metastable state $|\phi\rangle$ and
$\overline{{\it\Gamma}(t)} = \frac{1}{t}\int_{0}^{t}{\it\Gamma}(\varsigma)d\varsigma$ is the effective decay rate. Using (\ref{A1}) we get
\begin{equation}
{\cal P}(t) = e^{\textstyle{- \overline{\it\Gamma(t)}\,t}}. \label{P1}
\end{equation}

Important general properties of the amplitude ${\cal A}(t)$, and thus the survival probability ${\cal P}(t)$,  can be found using integral representation of ${\cal A}(t)$ as the Fourier transform.
Namely,
using a basis in ${\cal H}$ build from normalized eigenvectors $|E\rangle,\,\ E\in \sigma_{c}(\mathfrak{H}) = [E_{{min}}, {\infty})$ (where $\sigma_{c}(\mathfrak{H})$ is the
continuous part of the spectrum of $\mathfrak{H} $) of $\mathfrak{H}$ and using the expansion of $|\phi\rangle$ in this basis one can express the amplitude ${\cal A}(t)$ as the
following Fourier integral
\begin{equation}
{\cal A}(t) \equiv \int_{E_{min}}^{\infty} \omega(E)\,
e^{\textstyle{-\,\frac{i}{\hbar}\,E\,(t)}}\,d{E},
\label{a-spec}
\end{equation}
where $\omega(E) = \omega(E)^{\ast}$ and $\omega(E) > 0$ is the probability to find the energy of the system in the state $|\phi\rangle$ between $E$ and $E\,+\,dE$, and $E_{{min}}$ is
the minimal energy of the system. The last relation (\ref{a-spec}) means that the survival amplitude ${\cal A}(t)$ is a Fourier transform of an absolute integrable function $\omega
(E)$. If we apply the Riemann-Lebesgue lemma to the integral (\ref{a-spec}) then one concludes that there must be ${\cal A}(t) \to 0$ as $t \to \infty$. This property and the relation
(\ref{a-spec}) are the essence of the Fock--Krylov theory of unstable states \cite{Krylov:1947tmi,Fock:1978fqm}.
Within this approach the amplitude ${\cal A}(t)$, and thus the decay law ${\cal P}(t)$ of the metastable state $|\phi\rangle$, are determined completely by the density of the
energy distribution $\omega(E)$ for the system in this state \cite{Krylov:1947tmi,Fock:1978fqm} (see also \cite{Fonda,Kelkar:2010qn}, and so on). (This approach is also applicable to models in quantum field theory~\cite{Giacosa:2011xa,Giacosa:2018dzm}).

Assuming that $\langle \phi|\mathfrak{H}|\phi\rangle, \;\langle \phi|\mathfrak{H}^{2}|\phi\rangle$ exist and
that  \linebreak  $|\langle \phi|\mathfrak{H}|\phi\rangle| < \infty$ and $|\langle \phi|\mathfrak{H}^{2}|\phi\rangle| < \infty$ one can find that
(see \cite{Urbanowski:1994epq})
\begin{equation}
{\cal P}(t) \simeq 1 - \left(\frac{\delta \mathfrak{H}}{\hbar}\right)^{2}\,t^{2}, \label{P-0}
\end{equation}
at early times $t \to 0$, where $\delta \mathfrak{H} = \sqrt{\langle \phi|\mathfrak{H}^{2}|\phi
\rangle \,-\, E_{\phi}^{2}}$ and $E_{\phi} = \langle \phi|\mathfrak{H}|\phi\rangle$.
This formula can be obtained also, e. g., from the formula (\ref{amp})  using first terms of the
expansion of $ \exp\,[-\frac{i}{\hbar} \mathfrak{H} t]$ in series for $t \to 0$:
${\cal A}(t) \underset{t \to \, 0}{\simeq} 1 - \frac{i}{\hbar} \langle \phi|\mathfrak{H}|\phi\rangle -
 \frac{1}{2!}\left(\frac{t}{\hbar}\right)^{2}\langle \phi|\mathfrak{H}^{2}|\phi\rangle + \ldots$ .

Using (\ref{P-0}) and (\ref{G1}) one finds that at early times
\begin{equation}
\frac{{\it\Gamma}(t)}{\hbar} = 2 t \left(\frac{\delta \mathfrak{H}}{\hbar}\right)^{2}, \;\;({\rm for}\, \,t \to 0), \label{G-0}
\end{equation}
which agrees with the inference from Equations (\ref{h(t)}). Namely,
from the last term of Eq (\ref{h(t)}) one concludes that
\begin{equation}
h(t) \equiv E_{\phi} + \frac{\langle \phi|\mathfrak{H}\mathbb{Q}|\phi(t)\rangle}{{\cal A}(t)} \stackrel{\rm def}{=}  E_{\phi} + v(t). \label{h2}
\end{equation}
Taking into account that  $\mathbb{Q} = \mathbb{I} - \mathbb{P}$ and $\mathbb{P} = |\phi\rangle \langle \phi|$, and thus that $\mathbb{Q}|\phi\rangle = 0$,
the conclusion follows:
\begin{equation}
\begin{split}
h(0) \equiv \Re\,[h(0)]  &= E(0) \equiv E_{\phi}, \\
\Im\,[h(0)] &= 0\;\;\;\Rightarrow\; \;\;\;{\it\Gamma}(0) = 0, \label{hg=0}
\end{split}
\end{equation}
(if the matrix element $ \langle \phi |H| \phi\rangle$ exists).
On the other hand making use of the relation
$|\phi (t)\rangle =  \exp\,[-\frac{i}{\hbar} \mathfrak{H} t]|\phi\rangle$ one can find that at early times
\begin{equation}
v(t) \underset{t \to \, 0}{\simeq} - \frac{i}{\hbar}\langle \phi|\mathfrak{H}\mathbb{Q} \mathfrak{H}|\phi\rangle t
 \equiv -  \frac{i}{\hbar} \left(\delta \mathfrak{H}\right)^{2} t,   \label{v2}
\end{equation}
which leads to the same early time formula for ${\it\Gamma}(t)$ as that obtained directly from the definition (\ref{G1}) of ${\it\Gamma}(t)$.
It results from (\ref{h-m+g}), (\ref{h2}) and (\ref{v2})  that at early times
\begin{equation}
h(t) \underset{t \to \, 0}{\simeq} E_{\phi} -  \frac{i}{\hbar} \left(\delta \mathfrak{H}\right)^{2} t. \label{h+0}
\end{equation}
Hence
\begin{equation}
E(t) \underset{t \to \, 0}{\simeq} E_{\phi}, \;\;\;\;{\rm and}\;\;\;\;
{\it\Gamma}(t) \underset{t \to \, 0}{\simeq} \frac{2}{\hbar} \left(\delta \mathfrak{H}\right)^{2} t, \label{E+G-0}
\end{equation}
(compare Eq. (\ref{G-0})).
Note that the last relations are quite general and do not depend on the form of the function of the energy density  distribution $\omega(E)$ in the metastable state considered.

As it was shown  (see, e.g. \cite{Urbanowski:1994epq,Urbanowski:2016pks} and references therein), in a general case at later time when the decay law (the survival probability) has an exponential form to very good approximation, that is at {\em canonical decay times}, there are
\begin{equation}
\begin{split}
{\it\Gamma}(t) & \simeq {\it\Gamma}_{\phi}^{(1)}\stackrel{\rm def}{=} {\it\Gamma}_{0}, \;\;\;\;{\rm and},\\
E(t) & \simeq E_{0} \stackrel{\rm def}{=}  E_{\phi} -  \Delta_{\phi}^{(1)} \equiv E(0) - \Delta_{\phi}^{(1)},
\label{G11}
\end{split}
\end{equation}
where
\begin{equation}
{\it\Gamma}^{(1)}_{\phi} = 2\pi \langle \phi|\mathfrak{H}\mathbb{Q}\,\delta(\mathbb{Q}\mathfrak{H}\mathbb{Q} - E_{\phi})\,\mathbb{Q}\mathfrak{H}|\phi\rangle, \label{G3}
\end{equation}
is the decay width, and
\begin{equation}
\Delta^{(1)}_{\phi} = \langle \phi|\mathfrak{H}\mathbb{Q}\;\,{\rm P}_{v}\,\frac{1}{\mathbb{ Q}\mathfrak{H}\mathbb{Q} - E_{\phi} }\;\,\mathbb{Q}\mathfrak{H}|\phi\rangle, \label{Delta}
\end{equation}
(here ${\rm P}_{v}$ denotes principal value), and $\Delta^{(1)}_{\phi}$ is the correction to the energy of the system in the metastable state,
$\Delta_{\phi}^{(1)} = - \Re\,[v(t)]$, ($v(t) \simeq const.$ at canonical decay times)  and  $| \Delta_{\phi}^{(1)}| \ll | E_{\phi}|$.
  It should be noted that more detailed considerations show that these approximate results
 describe behavior of the system in the metastable state $|\phi\rangle$ accurate enough only for canonical decay times
 (i.e. when the exponential decay law holds with sufficient accuracy \cite{Khalfin}).

The representation of the survival amplitude ${\cal A}(t)$ as the Fourier transform (\ref{a-spec}) can be used to find the late time asymptotic form of ${\cal A}(t)$, ${\cal P}(t)$, the decay rate ${\it\Gamma}(t)$ and the instantaneous energy $E(t)$ (see \cite{Urbanowski:2006mw,Urbanowski:2008kra}). There are,
\begin{equation}
{{\it\Gamma}(t)\, \vline}_{\,t \rightarrow \infty} \stackrel{\rm def}{=} {\it\Gamma}_{lt}(t) =  \alpha_{1}\frac{\hbar}{t} + \alpha_{3} \left( \frac{\hbar}{t} \right)^{3} + \ldots, \label{g(t)-as}
\end{equation}
and
\begin{equation}
{E(t)\, \vline}_{\,t \rightarrow \infty} \stackrel{\rm def}{=} E_{lt}(t) = E_{min} + \alpha_{2}  \left(\frac{\hbar}{t}\right)^{2} + \alpha_{4} \left(\frac{\hbar}{t}\right)^{4} + \ldots, \label{E(t)-as}
\end{equation}
where  $\alpha_{k}$, are real numbers for $k=1,2, \ldots$ and $\alpha_{1} > 0$ and the sign of $\alpha_{k}$ for $k >1$ depends on the model considered (see
\cite{Urbanowski:2008kra}). Results  (\ref{g(t)-as}), (\ref{E(t)-as}) are completely general. They do not depend on the form of $\omega (E)$.
The absolute integrability of $\omega(E)$  and its first few derivatives are the only conditions that must be met to obtain the asymptotic expansions (\ref{g(t)-as}) and (\ref{E(t)-as}) (for details see \cite{Urbanowski:2008kra}).
An example of such a general function $\omega (E)$  is
$\omega (E) =  \omega_{\lambda} (E) =  {\it\Theta}(E-E_{min})\,(  E - { E}_{min})^{\lambda}\,\eta(E)$,
 where $\eta ({ E})$ is an absolutely integrable function,
 ${\it\Theta}(E)$ is a step function: ${\it\Theta}(E) = 0\;\;{\rm  for}\;\; E \leq 0$
and ${\it\Theta}(E) = 1\;\;{\rm for}\;\;E>0  $,
  and $ 0 \leq \lambda <1$, (see \cite{Urbanowski:2008kra,Urbanowski:2016pks}). In this case, in (\ref{g(t)-as}) we have   $\alpha_{1} = 2(\lambda +1)$, (see \cite{Urbanowski:2016pks}).

The above analysis shows that the following phases can be distinguished in the quantum decay process. The initial phase, $t \in [t_{0}^{init},T_{0})$, where $t_{0}^{init}$ is the instant of a creation of the quantum metastable state, and $T_{0}$ is the indicative maximal time such that ${\it\Gamma}(t)$ is smaller  than ${\it\Gamma}_{0}$ for $t < T_{0}$.
(For simplicity we assume that $t_{0}^{init} = 0$). In other words, the initial phase comprises a time period $t \in [0,T_{0})$ in which the approximations (\ref{G-0}), (\ref{E+G-0}) are sufficiently accurate and ${\it\Gamma}(t) < {\it\Gamma}_{0}$. The times $t \in (T_{0},T_{1})$ form the next phase called the {\em canonical decay} phase. During this phase, to a very high accuracy,
 the decay law has the form of an exponentially decreasing function of time. The time $T_{1}$ denotes approximately the end of the canonical decay phase and the beginning of the {\em  transition times} phase. For $ t > T_{1}$ the decay law as a function of time still decreases but slower and slower and begins to be oscillatory modulated up to the time $T_{2}$, when the late time  inverse power law phase begins to dominate. So times $t\in (T_{1},T_{2})$ form the transition times phase and times $t\in(T_{2},\infty)$ form the {\em late time inverse power law } phase.
 During the canonical decay phase, $t \in (T_{0},T_{1})$, the decay rate ${\it\Gamma}(t)$ and the instantaneous energy $E(t)$  are constant to a very good approximation and they are given by Eqs (\ref{G11}), (\ref{G3}), (\ref{Delta}).
  At times $t \in (T_{1},T_{2})$ the decay rate, ${\it\Gamma}(t)$, and the instantaneous energy, $E(t)$, are rapidly decreasing and oscillatory modulated functions of time. During the inverse power law phase, $t\in (T_{2}, \infty)$, the decay law behaves as a function of powers of $1/t$ and the decay rate ${\it\Gamma}(t)$ and the energy $E(t)$ take the form described by Eqs (\ref{g(t)-as}) and  (\ref{E(t)-as}) respectively.

In order to calculate the survival amplitude ${\cal A}(t)$ within
the Fock--Krylov theory of unstable states we need the energy density distribution function $\omega(E)$.
The simplest choice is
to take $\lambda = 0$ in formula for $\omega (E) \equiv \omega_{\lambda}(E)$
and to assume that  $\eta(E) $
has a Breit--Wigner (BW) form of the energy distribution density.
It turns out that the decay curves obtained in this simplest case are very similar in form to the curves calculated for
a more general $\omega (E)$,
(see \cite{Kelkar2021,Nowakowski} and analysis in \cite{Fonda}).
So to find the most typical properties of the decay process it is sufficient to make the relevant calculations for  $\omega (E)$ modeled by the the Breit--Wigner
distribution of the energy density:
\begin{equation}
\begin{split}
\omega (E) &\equiv \omega_{{BW}}(E)  \\
&\stackrel{\text{def}}{=} \frac{N}{2\pi}\, {\it\Theta} (E - E_{{min}}) \
\frac{{\it\Gamma}_{0}}{(E-E_{0})^{2} +
(\frac{{\it\Gamma}_{0}}{2})^{2}}, \label{omega-BW}
\end{split}
\end{equation}
where $N$ is a normalization constant.
The parameters $E_{0}$ and ${\it\Gamma}_{0}$ correspond to the energy of the system in the metastable state and its decay rate at the exponential (or canonical) regime of the decay
process. $E_{{min}}$ is the minimal (the lowest) energy of the system.
For  $\omega (E) = \omega_{BW}(E)$ one can find relatively easy an analytical form of ${\cal A}(t)$ (see e.g.  \cite{Sluis,R-U}).
Simply, inserting $\omega_{{BW}}(E)$ into formula (\ref{a-spec}) for the amplitude ${\cal A}(t)$ after some algebra one can find  ${\cal A}(t)$. Then
having the amplitude ${\cal A}(t)$ we can use it to
analyze properties of decay rate ${\it\Gamma} (t)$ and  the instantaneous energy $E(t)$.
These quantities are defined using the effective Hamiltonian $h(t)$ which is build from ${\cal A}(t)$. (Details of calculations can be found
 e. g. in \cite{ku2022,jcap-2020,Urbanowski:2006mw,Urbanowski:2008kra,Urbanowski:2009lpe,R-U}).
Then one can find
and  an  asymptotic analytical form of ${\cal A}(t), h(t)$, $E(t)$ and
${\it\Gamma} (t)$  at very late times (see e.g . \cite{Urbanowski:2006mw,Urbanowski:2009lpe,R-U}) as well as to find numerically the form of the survival probability ${\cal P}(t)$,  the decay rate ${\it\Gamma}(t)$ and the instantaneous energy $E(t)$ over
 a wide range of times $t$, starting from the initial instant $t=t_{0}^{init}=0$, through canonical decay and then intermediate times, up to asymptotically late times $t \to \infty$.
  So within the model based on the Breit--Wigner  distribution of the energy density one finds that at late times, $t \to \infty$,
there are
(see \cite{jcap-2020,R-U}):
\begin{equation} \label{Im-h-as}
{\it\Gamma}(t)  = - 2 \Im\,[h(t)]  \underset{t \to \, \infty}{\simeq}
2\,\frac{\hbar}{t}  + \ldots \, .
\end{equation}
and
\begin{equation}
 \label{Re-h-as}
E(t) = {\Re\,[h(t)]}\\
\underset{t \to \, \infty}{\simeq}
{ E}_{{min}}\, -\,2\,
\frac{ \beta }{{\it\Gamma}_{0}\,(\beta^{2} + \frac{1}{4}) } \,
\left(\frac{\hbar}{t} \right)^{2} + \ldots ,
\end{equation}
where  $\beta = \frac{E_{0} - E_{min}}{{\it\Gamma}_{0}} > 0$.

\begin{figure}[H]
 \begin{center}
\includegraphics[width=68mm]{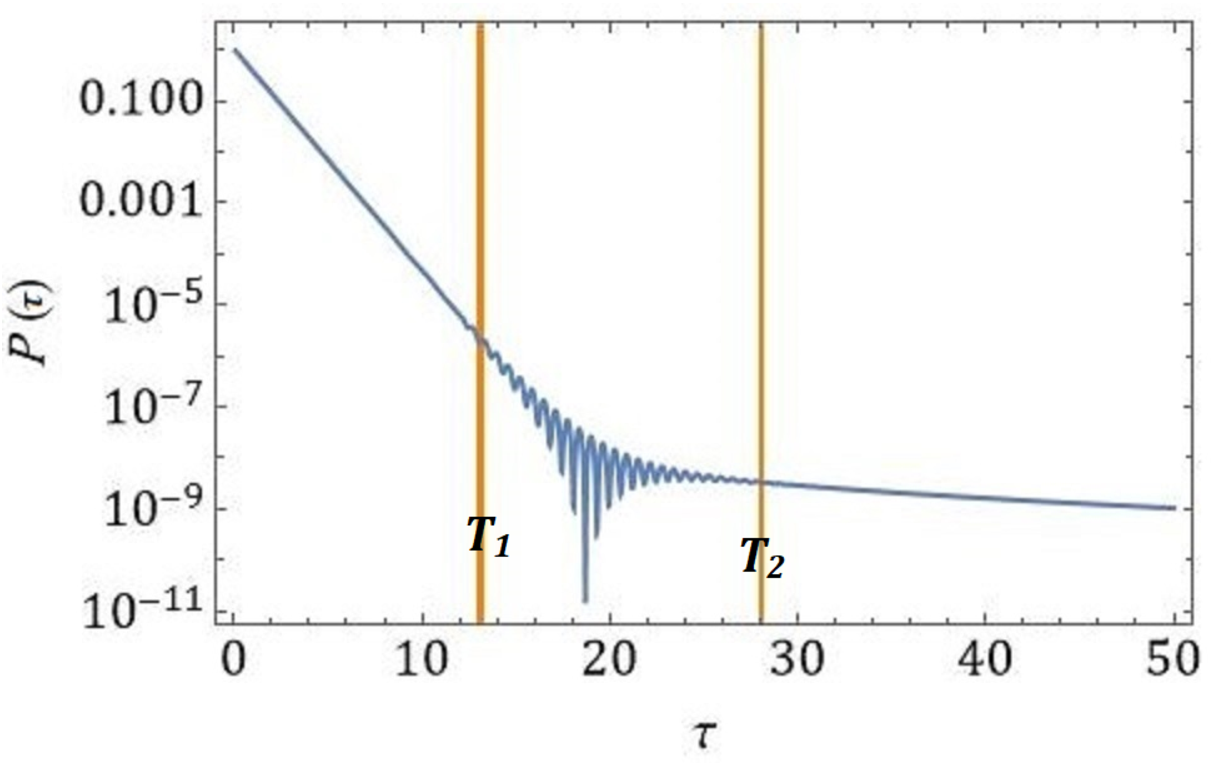}\\
\caption{A typical form of the survival probability ${\cal P}(\tau)$. Here $\tau  = \frac{t}{\tau_{0}}$.$\tau_{0} = \frac{\hbar}{{\it\Gamma}_{0}}$ is the lifetime, $t$ is the time. The case $\beta  = 10$.}
\label{p1}
 \end{center}
\end{figure}

\begin{figure}[H]
\begin{center}
\includegraphics[width=68mm]{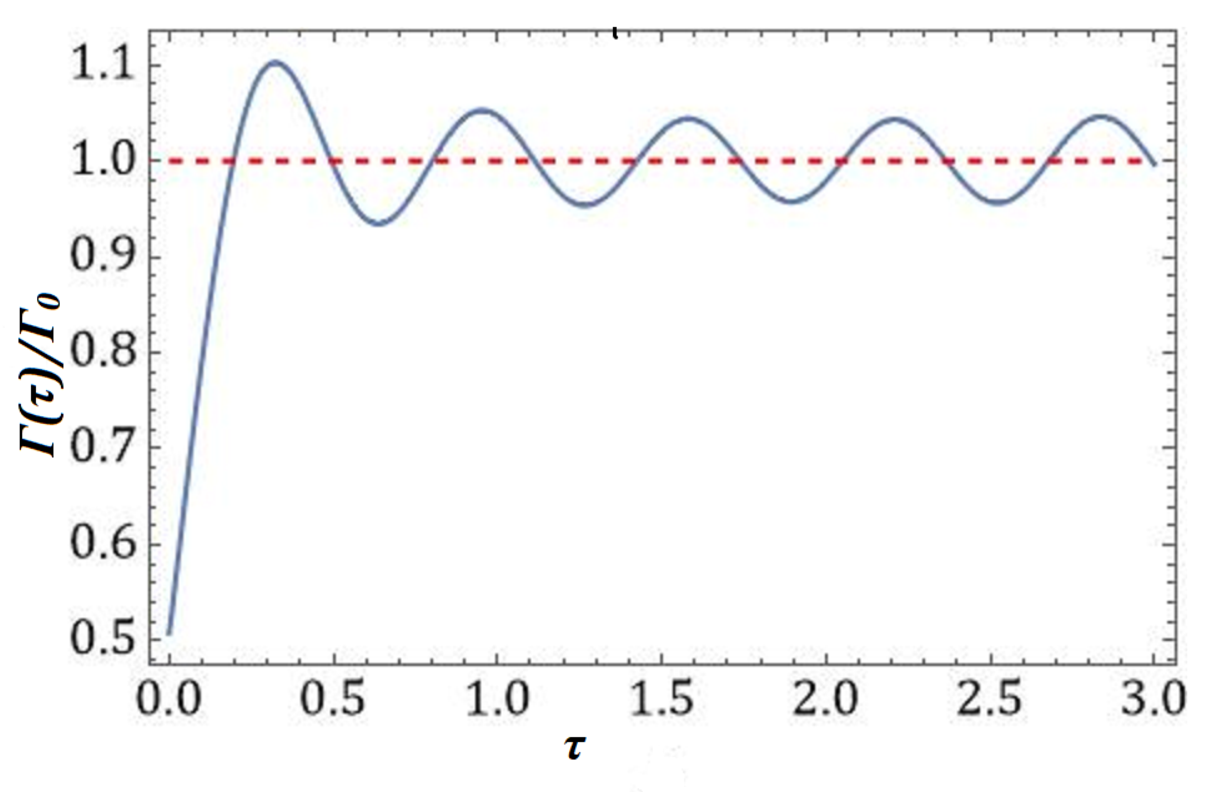}\\
\caption{The instantaneous decay rate ${\it\Gamma}(t)$ obtained for $\omega_{BW}(E)$ given by Eq. (\ref{omega-BW}).   The case $\beta  = 10$. $\tau  = \frac{t}{\tau_{0}}$,
$\tau_{0} = \frac{\hbar}{{\it\Gamma}_{0}}$ is the lifetime, $t$ is the time.  The solid line:  The ratio ${\it\Gamma}(t)/{\it\Gamma}_{0}$.  The dashed line:  ${\it\Gamma}(t) = {\it\Gamma}_{0}$.}
  \label{g1}
\end{center}
\end{figure}

\begin{figure}[H]
\begin{center}
\includegraphics[width=68mm]{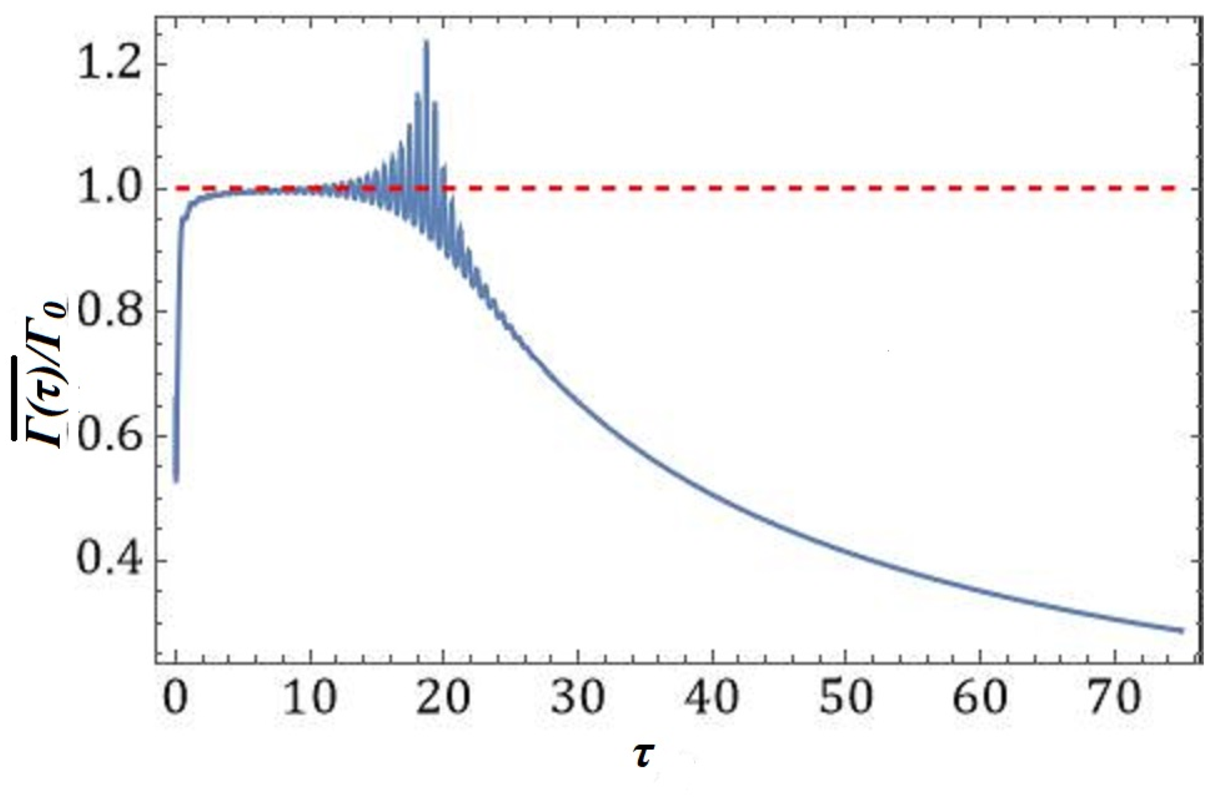}\\
\caption{The effective decay rate $\overline{{\it\Gamma}(t)}\equiv \frac{1}{t}\int_{0}^{t}{\it\Gamma}(\varsigma)\,d\varsigma$ obtained for $\omega_{BW}(E)$ given by Eq. (\ref{omega-BW}).   The case $\beta  = 10$. $\tau  = \frac{t}{\tau_{0}}$,
$\tau_{0} = \frac{\hbar}{{\it\Gamma}_{0}}$ is the lifetime, $t$ is the time.  The solid line:  The ratio $\overline{{\it\Gamma}(t)}/{\it\Gamma}_{0}$.  The dashed line:  $\overline{{\it\Gamma}(t)}= {\it\Gamma}_{0}$.
}
  \label{ge2}
\end{center}
\end{figure}

\begin{figure}[H]
\begin{center}
\includegraphics[width=68mm]{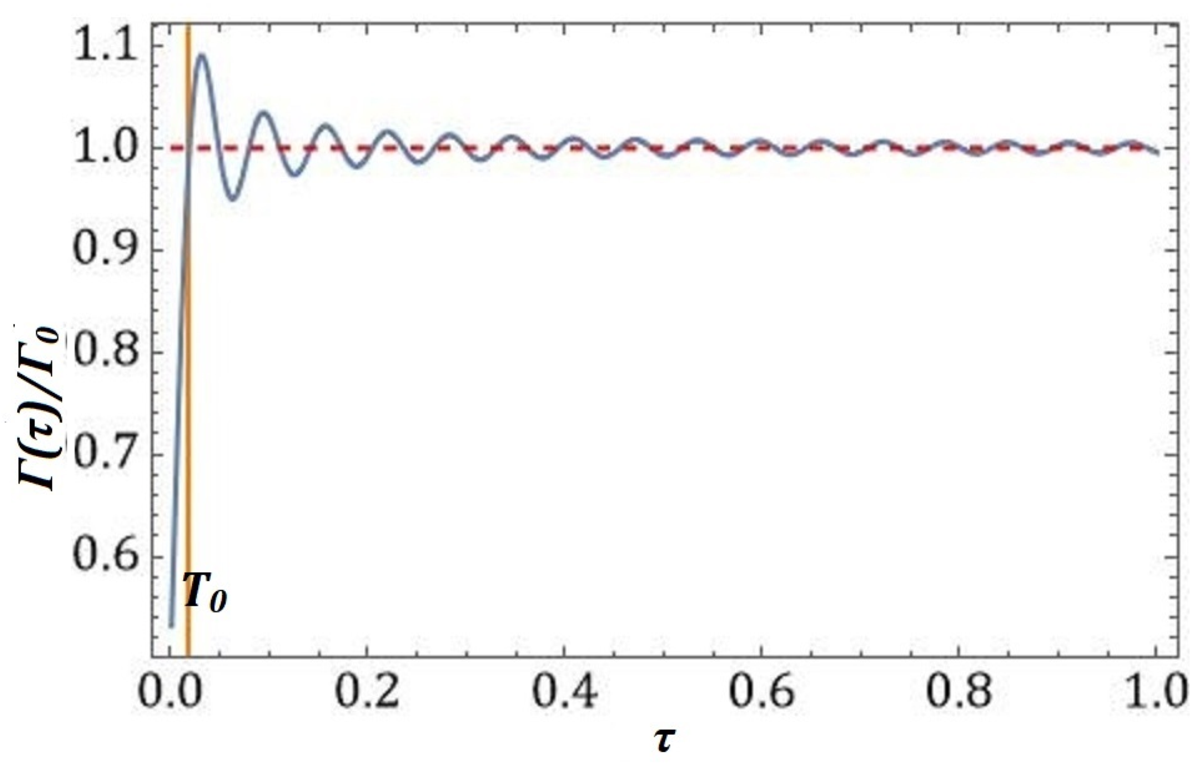}\\
\caption{The instantaneous decay rate ${\it\Gamma}(t)$ obtained for $\omega_{BW}(E)$ given by Eq. (\ref{omega-BW}).   The case $\beta  = 100$. $\tau  = \frac{t}{\tau_{0}}$,
$\tau_{0} = \frac{\hbar}{{\it\Gamma}_{0}}$ is the lifetime, $t$ is the time.  The solid line:  The ratio ${\it\Gamma}(t)/{\it\Gamma}_{0}$. The dashed line:
${\it\Gamma}(t) = {\it\Gamma}_{0}$.}
  \label{g3}
\end{center}
\end{figure}

  Results of the numerical calculations depend on the value of the ratio $\beta$ (see \cite{ku2022} for details).
  A visualization of these results is presented below in graphical form in Figs \ref{p1}, \ref{g1}, \ref{ge2}, \ref{g3}, \ref{e1}, \ref{e2}, \ref{k5}.  There is
\begin{equation}
\kappa (t)  = \frac{E^{\,\text{F}}(t) - E_{{min}}}{E_{0}^{\,\text{F}} - E_{{min}}}, \label{kappa}
\end{equation}
in Figs \ref{e1},  \ref{e2} and \ref{k5}.
In some of these Figures, the aforementioned times $T_{0}, T_{1}$ and $T_{2}$ are marked for illustration.
And so,
the time $T_{0}$
denotes the maximal time such that ${\it\Gamma}(t)$ is smaller then
 ${\it\Gamma}_{0}$ for $t < T_{0}$:
There is ${\it\Gamma}(t) < {\it\Gamma}_{0}$ for $t \in [0,T_{0})$
(see Fig \ref{g3}).
The time $T_{1}$
is the time when the canonical (exponential) decay law ceases to apply and the transition from the exponential decay phase to the phase in which the decay law takes the form of powers $1/t$ begins, simply $T_{1}$ denotes approximately the beginning of the transition times
(see Figs \ref{p1},  \ref{e1}, \ref{e2}).
The time $T_{2}$
denotes an indicative end of the transition phase between exponential and inverse power like forms of the decay law (it is also the indicative beginning of this last phase
 --- see Fig \ref{e2}). In general  the transition times are such times $t$ that $t \in (T_{1},T_{2})$.

\section{Metastable states: cosmological applications}

Within the standard approach  transitions from a meta\-stable false vacuum  to the true vacuum state  are  described in cosmology using quantum field theory methods (see eg.
\cite{Coleman:1977py} --- \cite{Stojk} and many other papers).
If one wants to generalize the  results obtained in the previous Section on the basis of quantum mechanics to quantum field theory one should take into account among others  volume factors so that
survival probabilities per unit volume  should be considered and similarly the energies and the decay rate: $E \mapsto  \varepsilon  = \frac{E}{V_{0}}$, ${\it\Gamma}_{0} \mapsto \gamma =
\frac{{\it\Gamma}_{0}}{V_{0}}$, where $V_{0} = V(t_{0}^{init})$ is the volume of the considered system at the initial instant $t_{0}^{init}$, when the time evolution starts.
The volume $V_{0}$ is used in these considerations because
the initial unstable state $|\phi\rangle \equiv |0\rangle^{\text{F}}$ at $t=t_{0}^{init}=0$ is expanded into eigenvectors $|E\rangle$ of $\mathfrak{H}$,
 (where $E \in \sigma_{c}(\mathfrak{H})$),  and then this expansion is used to find the density of the energy distribution $\omega (E)$  at this initial instant $t_{0}^{init}$.
Now, if we identify $\varepsilon_{{de}}(t_{0}^{init})$ with the energy $E_{0}^{\;\text{F}}$ of the unstable system divided by the volume $V_{0}$:
$\varepsilon_{{de}}(t_{0}^{init})  \equiv \varepsilon_{0}^{\,\text{F}} \equiv \varepsilon_{0}^{{qft}} \stackrel{\rm def}{=} \varepsilon_{{de}}^{0} = \frac{E_{0}^{\,\text{F}}}{V_{0}}$ and $ \varepsilon_{{bare}}
=\frac{E_{{min}}}{V_{0}}$, (where $\varepsilon_{{0}}^{{qft}}$ is the vacuum energy density calculated using quantum field theory methods) then
it is easy to see that the mentioned changes
$E \mapsto  \frac{E}{V_{0}}$ and  ${\it\Gamma}_{0} \mapsto \frac{{\it\Gamma}_{0}}{V_{0}}$ do not changes the parameter $\beta$:
\begin{equation}
\beta = \frac{E_{0}^{\,\text{F}} - E_{{min}}}{{\it\Gamma}_{0}} \equiv  \frac{\varepsilon_{de}^{0} - \varepsilon_{bare}}{{\gamma}_{0}} > 0, \label{beta-rho}
\end{equation}
(where $\gamma_{0} = {\it\Gamma}_{0}/V_{0}$, or equivalently, $ {\it\Gamma}_{0}/V_{0} \equiv \frac{\varepsilon_{de}^{0} - \varepsilon_{bare}}{\beta}$).
This means that the relations (\ref{E(t)-as}),  (\ref{Re-h-as}),
can be replaced by
corresponding relations for the densities $\varepsilon_{{de}}$ or $\Lambda$ (see, eg., \cite{jcap-2020,Urbanowski:2016pks,Urbanowski:2012pka,ms-ku2}).
Simply, within this approach $E(t)=E^{\,\text{F}}(t)$ corresponds to the running cosmological constant $\Lambda(t)$ and $E_{{min}}$ to the $\Lambda_{bare}$. For example,  we have
\begin{equation}
\begin{split}
\kappa (t) & = \frac{E^{\,\text{F}}(t) - E_{{min}}}{E_{0}^{\,\text{F}} - E_{{min}}}
\equiv \frac{\frac{E^{\,\text{F}}(t)}{V_{0}} - \frac{E_{{min}}}{V_{0}}}{\frac{E_{0}^{\,\text{F}}}{V_{0}} - \frac{E_{{min}}}{V_{0}}} \\
&= \frac{\varepsilon^{F}(t) - \varepsilon_{bare}}{\varepsilon^{F}_{0} - \varepsilon_{{bare}}}
= \frac{\varepsilon_{{de}}(t) - \varepsilon_{{bare}}}{\varepsilon_{{de}}^{0} - \varepsilon_{{bare}}} \\ & = \frac{\Lambda (t) - \Lambda_{{bare}}}{\Lambda_{0} - \Lambda_{{bare}}}.
\label{kappa-L}
\end{split}
\end{equation}
Here
$\varepsilon^{F}(t) =  \frac{E^{\,\text{F}}(t)}{V_{0}}\equiv \varepsilon_{{de}}(t) \equiv \rho_{de}c^{2}$,  $\Lambda (t) = \frac{8\pi G}{c^{4}}\,\varepsilon_{de}(t) \equiv
\frac{8\pi G}{c^{2}}\,\rho_{de}(t) $.
Equivalently, $\rho_{de}(t) = \frac{c^{2}}{8\pi G} \Lambda(t)$.

\begin{figure}[h!]
\begin{center}
\includegraphics[width=68mm]{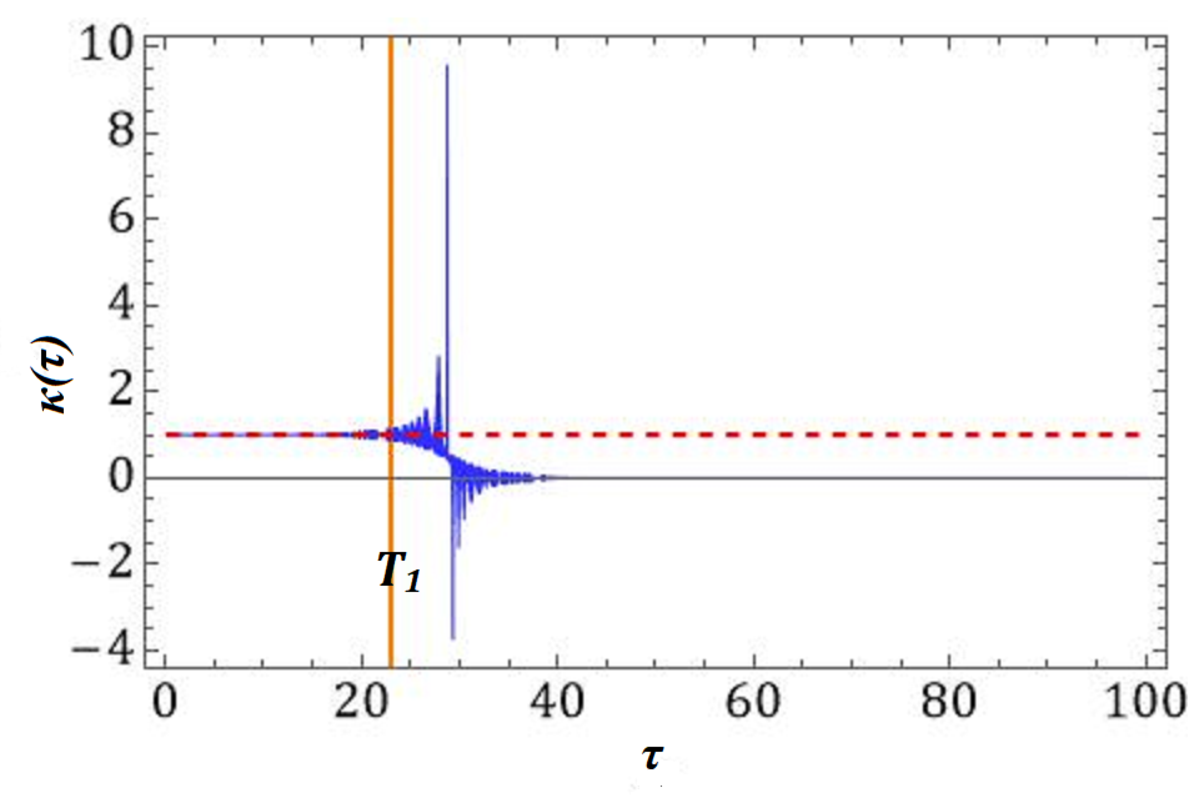}\\
\caption{
 An illustration of a behavior of energy $E(\tau)$ (or cosmological "constant" $\Lambda(t)$).
Results obtained for  $\omega_{BW}(E)$ given by Eq. (\ref{omega-BW}). The case $\beta =100$. The solid line: $\kappa (\tau ) = \left(E(\tau) - E_{min}\right)/\left(E_{0} - E_{min}\right) \equiv \left( \Lambda (t) - \Lambda_{bare}\right)/\left(\Lambda_{0} - \Lambda_{bare}\right)$.  The dashed line: $E(\tau) = E_{0} = \text{const}$,  or,  $\Lambda (\tau) = \Lambda_{0} = \text{const}$, ($\kappa (\tau) = 1$).
The time, $t$,  is measured in lifetimes $\tau_{0}$:  $\tau  =t/\tau_{0} $ and    $\tau_{0} = \hbar/{\it\Gamma}_{0}$ is the lifetime.
}
  \label{e1}
\end{center}
\end{figure}

\begin{figure}[h!]
\begin{center}
\includegraphics[width=68mm]{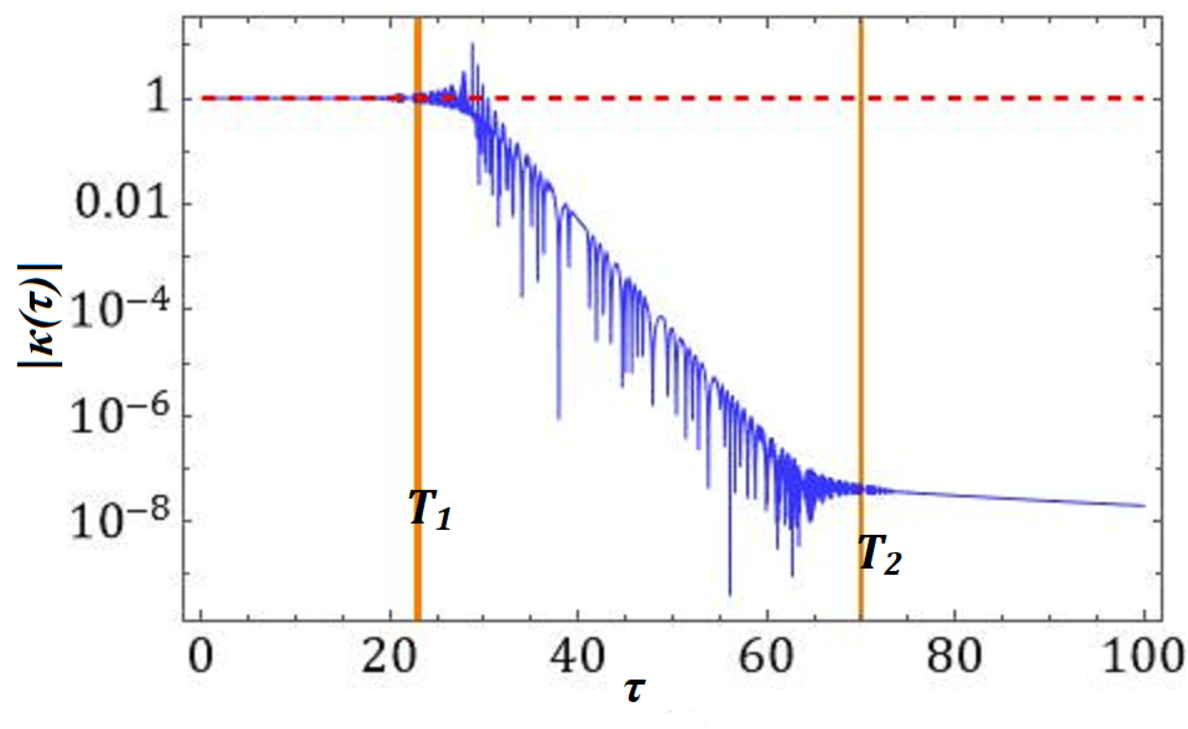}\\
\caption{
The modulus of the results presented in Fig (\ref{e1}) in the logarithmic scale. The solid line:  $\left|\kappa (\tau )\right| = \left|\left(E(\tau) - E_{min}\right)/\left(E_{0} - E_{min}\right)\right| \equiv \left|\left( \Lambda (t) - \Lambda_{bare}\right)/\left(\Lambda_{0} - \Lambda_{bare}\right)\right|$.  The dashed line: $E(\tau) = E_{0} = \text{const}$, or,  $\Lambda (\tau) = \Lambda_{0} = \text{const}$, ($\kappa (\tau) = 1$).
The time, $t$,  is measured in lifetimes $\tau_{0}$:  $\tau  =t /
\tau_{0} $ and    $\tau_{0} = \hbar/{\it\Gamma}_{0}$ is the lifetime.
}
  \label{e2}
\end{center}
\end{figure}

Our analysis shows that the properties of the instantaneous energy $E(t)$ change with time. The same is true for $\Lambda (t)$ (see Figs \ref{e1}, \ref{e2}).
Taking into account relations connecting $E(t)$ and $\Lambda (t)$ and using  (\ref{E+G-0}), (\ref{G11}) and (\ref{E(t)-as})
one can conclude that
there should be,
\begin{equation}
\begin{split}
\Lambda (t) &\simeq \Lambda_{0} \simeq \frac{8\pi G}{c^{4}}\,\varepsilon_{0}^{F} \\
&\equiv
\frac{8\pi G}{c^{4}}\,\frac{ ^{\text{F}}\langle 0 |\mathfrak{H}| 0 \rangle^{\text{F}} }{V_{0}}, \;\;\; t\in [0,T_{0}) \cup (T_{0},T_{1}), \label{L-infl}
\end{split}
\end{equation}
at canonical decay times $t < T_{1}$, were $\varepsilon_{0}^{F} = \varepsilon_{0}^{qft} = \frac{^{\text{F}}\langle 0 |\mathfrak{H}| 0 \rangle^{\text{F}}}{V_{0}}$.  (Here we used relations (\ref{E+G-0}), (\ref{G11})  and  the property that $|\Delta_{\phi}^{(1)}| \ll |E_{\phi}|$ from which it follows that in our analysis it
is enough to assume that
$E_{0}^{\,\text{F}} \simeq E_{\phi}$, i.e., that $E_{0}^{\,\text{F}} \simeq\;\,  ^{\text{F}}\langle 0 |\mathfrak{H}| 0 \rangle^{\text{F}}$). In other words there should be $\Lambda
(t) \simeq \Lambda_{0} \equiv \Lambda_{qft} = \frac{8\pi G}{c^{4}}\,\varepsilon_{0}^{qft}$ at times $t < T_{1}$.
Then latter, when time $t$ runs from  $t =T_{1}$ to  $t = T_{2}$ the quantum
effect discussed above forces this $\Lambda(t) \simeq \Lambda_{0}$ to reduce its value  for  $(t > T_{2})$ to the following one:
\begin{equation}
\Lambda(t) \simeq \Lambda_{eff}(t) = \Lambda_{\text{bare}} + \frac{\alpha_{2}^{\Lambda}}{t^{2}} + \frac{\alpha_{4}^{\Lambda}}{t^{4}} + \ldots \ll \Lambda_{0},\; (t > T_{2}).
  \label{lambda3}
\end{equation}
(A detailed discussion of this case is given in \cite{jcap-2020,Urbanowski:2012pka,ms-ku2,epjc-2017b}).
At times $t \in (T_{1}, T_{2})$ the energy $E(t)$ is oscillatory modulated and decreases from the value $E(t) = E_{0}$ to the late time asymptotic value $E_{lt}(t)$ (see (\ref{E(t)-as}).  An analogous behavior takes place for  $\Lambda (t)$ at this time interval. The typical behavior of  $\Lambda(t)$ as a function of time is shown
below
in Figs \ref{e1}, \ref{e2}, \ref{k5}.

\begin{figure}[h!]
\begin{center}
\includegraphics[width=68mm]{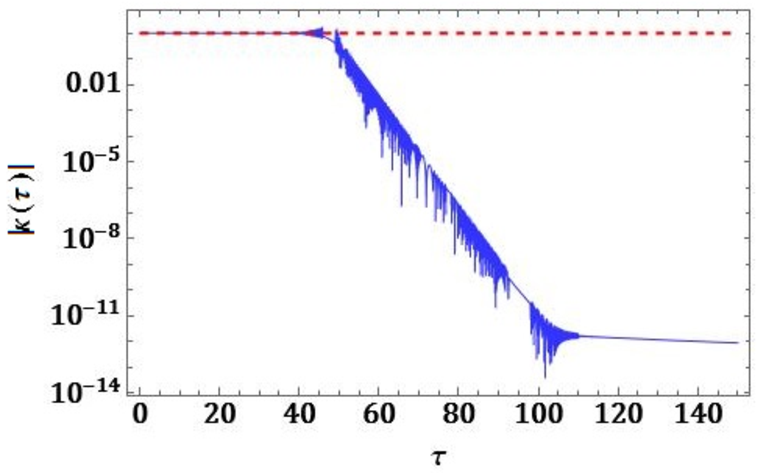}\\
\caption{
The modulus of $\kappa(\tau)$ in the logarithmic scale.  The case  $\beta = 10000$. The solid line:  $\left|\kappa (\tau )\right| = \left|\left(E(\tau) - E_{min}\right)/\left(E_{0} - E_{min}\right)\right| \equiv \left|\left( \Lambda (t) - \Lambda_{bare}\right)/\left(\Lambda_{0} - \Lambda_{bare}\right)\right|$.  The dashed line: $E(\tau) = E_{0} = \text{const}$, or,  $\Lambda (\tau) = \Lambda_{0} = \text{const}$,  ($\kappa (\tau) = 1$).
The time, $t$,  is measured in lifetimes $\tau_{0}$:  $\tau  =t /
\tau_{0} $ and    $\tau_{0} = \hbar/{\it\Gamma}_{0}$ is the lifetime.
}
  \label{k5}
\end{center}
\end{figure}

\section{Analysis and discussion}

We will start our analysis and discussion with the assumption that $E_{0}^{F}> 0$, so that $\varepsilon_{0}^{qft} \equiv \varepsilon_{0}^{F} >0$  and therefore $\Lambda_{qft} = \Lambda_{0} > 0$.
According to conclusion drawn in Sec. 2 the universe born in a false vacuum state will survive well up to  late times if
the rate $\nu_{V}(t) $ of the expansion of the effective volume $V_{eff}(t)$ is nonnegative. Using the formula  (\ref{d-r1}) for  $\nu_{V}(t)$ this conclusion can be written as follows
\begin{equation}
\nu_{V}(t) = 3\frac{\dot{a}(t)}{a(t)}\, - \frac{{\it\Gamma}(t)}{\hbar}\geq  0, \label{s1}
\end{equation}
or, equivalently,
\begin{equation}
3H(t)\, - \frac{{\it\Gamma}(t)}{\hbar} \geq  0, \label{s2}
\end{equation}
where $H(t) = \frac{\dot{a}(t)}{a(t)}$ is the Hubble parameter and ${\it\Gamma}(t) $ is the decay rate of the metastable false vacuum state. Using  the first Einstein equation
(\ref{Fr1})
one can replace the condition (\ref{s1}) by the following one:
\begin{equation}
\left(\frac{{\it\Gamma}(t)}{\hbar}\right)^{2}  \leq 9\left[ - \frac{kc^{2}}{R_{0}^{2}\,a^{2}(t)} +  \frac{8\pi G_{N}}{3}\,\rho +\frac{\Lambda\,c^{2}}{3} \right].  \label{s3}
\end{equation}
Note that inequalities (\ref{s2}), (\ref{s3}) work also in the case of a stable vacuum. Simply in such a case we have ${\it\Gamma}(t) \equiv 0$ and these inequalities are fulfilled trivially.

Now if the system is in the false vacuum state, $|0\rangle^{F}$,  then at times $t \in [0,T_{0})$, (where $t_{0}^{init} = 0$ is the initial instant), and
at canonical decay times,  $T_{0} < t < T_{1}$, the energy of the system in this state equals
$E^{\,\text{F}}(t) = E_{0}^{\,\text{F}} \approx \,  {^{F}\langle} 0|\mathfrak{H}|0\rangle^{F}$ to a very good approximation and then $\Lambda (t) \simeq \Lambda_{0}$ is given by the formula (\ref{L-infl}) and it can be very large.

At very early times, $t \in [0,T_{0})$, there is no matter  and the vacuum energy dominates. Therefore the behavior of expansion rate $\dot{a}(t)$ at these times is such that the curvature signature in Eq (\ref{Fr1}) can be approximated as  $ k \approx 0$ (see, e.g. \cite{Cheng}) and then Eq (\ref{Fr1}) simplifies to
\begin{equation}
H^{2}(t) = \frac{{\dot{a}}^{2}(t)}{a^{2}(t)}   \simeq \frac{\Lambda_{0}\,c^{2}}{3}.  \label{Fr1a}
\end{equation}
So in this case the universe considered will survive well at these times if the following inequality replacing (\ref{s3}) will be satisfied:
\begin{equation}
\left(\frac{{\it\Gamma}(t)}{\hbar}\right)^{2} \leq 3 \Lambda_{0}\,c^{2} \simeq   24\pi G_{N}\,\frac{^{\text{F}}\langle 0|\mathfrak{H}|0\rangle^{\text{F}}}{c^{2}\,V_{0}} \equiv
 24\pi G\,\frac{\varepsilon^{F}_{0}}{c^{2}}. \label{s4}
\end{equation}
Taking into account the results (\ref{G-0}), (\ref{E+G-0}), i.e. that,  $\frac{{\it\Gamma}(t)}{\hbar} \underset{t \to \, 0}{\sim} t$, (see also Figs \ref{g1}, \ref{g3}) one can conclude that at very early times, $t \in [0,T_{0})$, the left hand side of the inequality (\ref{s4}) is always smaller then the right hand side of this inequality. This means that  at these times the universe can not decay.

Recent studies based on the results of astronomical observations suggest that our Universe is flat \cite{Ef,Ai,Ag}. Therefore if one wants to apply Eqs. (\ref{s2}), (\ref{s3}) to our Universe one should insert $k=0$ for the curvature $k$ in Eq. (\ref{Fr1}). Then the first Friedmann Equation simplifies to the following one:
\begin{equation}
H^{2}(t)
= \frac{8\pi G_{N}}{3}\,\rho +\frac{\Lambda\,c^{2}}{3}, \label{Fr3}
\end{equation}
which allows to rewrite the condition  (\ref{s3}) as follows
\begin{equation}
 \left(\frac{{\it\Gamma}(t)}{\hbar}\right)^{2} \leq 3 \left[ 8\pi G_{N} \,\rho\, +\, \Lambda\,c^{2} \right], \label{s5}
\end{equation}
and only this case will be analyzed in the following.

Let us assume now that the lifetime, $\tau_{0}^{F}$,  of the false vacuum state is much shorter then duration of the inflationary epoch and that the time $T_{1}$ is comparable with the duration of this epoch (such a scenario was analyzed in \cite{ku2022}).
Then
let us  consider times $t \in (T_{0},T_{1})$, i. e. the epoch of the canonical decay times. In such a case, similarly to the case of very early times $t \in [0,T_{0})$, we can ignore the matter density $\rho$ in Eqs (\ref{Fr1}), (\ref{Fr3}) and the curvature $k$ in Eq (\ref{Fr1}) can be approximated as $ k \approx 0$.
So in this case the universe considered will survive well at  times $T_{0} < t < T_{1}$ if again the inequality (\ref{s4})  will be satisfied.

 Note that at the canonical decay times (see (\ref{G3}) and Figs \ref{g1}, \ref{ge2}, \ref{g3}) there is ${\it\Gamma}(t) \simeq {\it\Gamma}^{(1)} \simeq {\it\Gamma}_{0}$ to a very good approximation. So considering the false vacuum
we can replace ${\it\Gamma}(t)$ in (\ref{s4}) by ${\it\Gamma}_{0}$ and use ${\it\Gamma}_{0} \equiv {\it\Gamma}_{F}$ to obtain
\begin{equation}
\left( \frac{{\it\Gamma}_{F}}{\hbar} \right)^{2} \equiv \left(\frac{1}{\tau_{0}^{F}}\right)^{2} \leq
 24\pi G_{N}\,\frac{\varepsilon^{F}_{0}}{c^{2}}, \label{s4a}
\end{equation}
for the epoch considered,  $t\in (T_{0}, T_{1})$.  Here $\tau_{0}^{F} = \frac{\hbar}{{\it\Gamma}_{F}}$ and ${\it\Gamma}_{F}$ are  the lifetime and  decay rate (decay width) of the system in the metastable false vacuum state $|0\rangle^{F}$  respectively.
There is $\frac{24\pi G_{N}}{c^{2}} \simeq 55,9 \times 10^{-27}$ [m/kg] and hence $\frac{c^{2}}{24\pi G_{N}} \simeq 1.8 \times 10^{25}$ [kg/m].
So, for the epoch $t\in (T_{0}, T_{1})$ the condition (\ref{s4a}) can be rewritten as follows:
\begin{equation}
\varepsilon^{F}_{0} \geq 1.8 \times 10^{25}{\,\rm [kg/m]} \times \left(\frac{1}{\tau_{0}^{F}}\right)^{2}. \label{s4b}
\end{equation}

As it has been stated in Sec. 1, this paper complements  analysis presented in \cite{ku2022}, where the possibility that vacuum energy could drive  the inflation process (or contribute significantly to it) was discussed. Within such a scenario, as an example, it was analyzed the case that $T_{1} \sim (50 - 100) \tau_{0}^{F}$ and $\tau_{0}^{F} \sim (10^{-38} - 10^{-36})$[s].
We will begin our analysis of the possibility that the Universe will survive with this scenario.
 So,
 let us consider the case $\tau_{0}^{F} \sim 10^{-36}$ [s]: From (\ref{s4b}) if follows that in such a case
the Universe can survive at the epoch $t \in (T_{0}, T_{1})$ only if
$\varepsilon_{0}^{F} > 1.8 \times 10^{97}$ $[\frac{{\rm J}}{{\rm m}^{3}}]$.
Similarly,
if to assume that $\tau_{0}^{F} \sim 10^{-38}$ [s], then inequality (\ref{s4b}) implies that the Universe has a chance of the survival only when
$\varepsilon_{0}^{F} > 1.8 \times 10^{101}$  $[\frac{{\rm J}}{{\rm m}^{3}}]$. The surprise in this scenario is that the minimum values of the false vacuum energy density, $\varepsilon_{0}^{F}$,  thus obtained fit well with the energy density resulting from the estimates made on the assumption that quantum field theory can be applied up to the Planck scale, $M_{Pl} \sim
10^{19}$ [GeV/$c^{2}$], (see e.g. Sec. 7 in \cite{Sahni}, Eq. (19) in \cite{Carroll},  or  the end of Sec. 3.1 in \cite{Rugh}).

In order to analyze  inequalities (\ref{s2}) or (\ref{s5}) for the  epoch $t \in (T_{1},T_{2})$ within this scenario one should know the form of the potential $V(\Phi)$ and thus the Lagrangian and the Hamiltonian $\mathfrak{H}$ allowing to find $|0\rangle^{F}$ and then to calculate ${\it\Gamma} (t)$ and $E(t)$ for $t \in (T_{1},T_{2})$. It is because  ${\it\Gamma} (t)$ is scillatory modulated in this epoch and these modulations depend on $\mathfrak{H}$. The same is true for  $E(t)$ and therefore for  $\Lambda(t)$ (see Figs \ref{e1}, \ref{e2} and \ref{k5}).

Within the scenario analyzed the picture is simpler at the epoch $t\in (T_{2},\infty)$.
At this epoch ${\it\Gamma} (t)$ is given by the formulae (\ref{g(t)-as}), (\ref{Im-h-as}) and
(\ref{lambda3}). (Note that within the scenario analyzed we are living now at the epoch $t > T_{2}$, i.e. $t\in (T_{2},\infty)$).  This means that now
we should use Eq. (\ref{Im-h-as}).
Inserting ${\it\Gamma}(t)   \underset{t \to \, \infty}{\simeq}
2\,\frac{\hbar}{t}$ into (\ref{s2})
we obtain the following inequality
\begin{equation}
\frac{2}{t} \leq 3 H(t), \label{s-T2a}
\end{equation}
or
\begin{equation}
\frac{3}{2} \,t\,H(t) \geq 1, \label{s-T2b}
\end{equation}
for $t \in (T_{2}, \infty)$.
First, let us consider for simplicity the  special case, assuming for a moment that
that $t \sim t_{0}$, where $t_{0}$ is the age of the Universe today.  Then using the present--day values of $H(t) = H(t_{0}) = H_{0}$ and $t_{0}$ (see \cite{pdg-2020}) one obtains that $\frac{3}{2} \,t_{0}\,H_{0} \simeq 1.4$. So the condition (\ref{s-T2b}) holds for the today  Universe and today we are safe.
In general, if the universe has successfully survived the epoch $(T_{1},T_{2})$  and
has been  fortunate enough to survive until
$t > T_{2}$, then
the simplest choice is to use solutions of the Friedmann Eqs. (\ref{Fr2}), (\ref{Fr3}),
\begin{equation}
H(t) = \frac{2}{3t}, \label{H-mdu}
\end{equation}
for the matter dominated universe (see, e.g. \cite{Cheng}). Inserting (\ref{H-mdu}) into the weak inequality (\ref{s-T2b}) one gets the result $\frac{3}{2} \,t\,H(t) = 1$ for $t\in(T_{2},\infty)$. So in such a case the universe should not decay.

Note that the above analysis concerning the epoch $t \in (T_{2},\infty)$ bases on the assumption that a metastable state is modeled by the Breit--Wigner energy density distribution which leads, among others, to the result (\ref{Im-h-as}) used in Eqs. (\ref{s-T2a}), (\ref{s-T2b}). In the general case considering the epoch $t\in (T_{2},\infty)$ one should use Eq. (\ref{g(t)-as}) instead of (\ref{Im-h-as}), which leads to the following inequality,
\begin{equation}
\frac{\alpha_{1}}{t} \leq 3 H(t), \label{s-T2gen}
\end{equation}
replacing (\ref{s-T2a}), where $\alpha_{1}$ depends on the model considered. So, depending on $\alpha_{1}$ and $H(t)$, (or on the right hand side  of (\ref{s5})), the inequality
can be satisfied or not.

Let us consider now another scenario. Let
$\tau_{0}^{F}  \sim t_{0}$.   Note that if this assumption is true, then  ${\it\Gamma}(t) \simeq {\it\Gamma}_{F}$ and $T_{1} > t_{0}$ and are living
in the canonical decay times epoch
$t \in (T_{0}, T_{1})$. As a result the term $\frac{{\it\Gamma}(t)}{\hbar}$ in (\ref{s2}) can be replaced by $ \frac{{\it\Gamma}(t)}{\hbar} = \frac{{\it\Gamma}_{F}}{\hbar} = \frac{1}{\tau_{0}^{F}}  \simeq \frac{1}{t_{0}}$. Hence the inequality (\ref{s2}) can be rewritten as follows
\begin{equation}
3 \,\tau_{0}^{F}\, H(t) \equiv 3 \,t_{0} \,H(t_{0}) \geq 1. \label{sTu}
\end{equation}
Taking $t\sim t_{0}$ and using the present--day values of $H(t_{0}) = H_{0}$ and $t_{0}$ one finds that
$3 \,t_{0} \,H_{0} \simeq  2,798$. This means   that
when $\tau_{0}^{F}  \sim t_{0}$,  then  the condition (\ref{sTu}) is satisfied and in this case our Universe should survive well up to times $t > T_{1} > t_{0}$. Unfortunately another problem arises. Namely if the lifetime is of order of the age of the Universe and $t < T_{1}$ then, according to Eg. (\ref{L-infl}), $\Lambda (t) = \Lambda_{0} \simeq \frac{8\pi G}{c^{4}}\,\varepsilon_{0}^{F}\simeq const.$, where $\varepsilon_{0}^{F} = \varepsilon_{0}^{qft} = \frac{^{\text{F}}\langle 0 |\mathfrak{H}| 0 \rangle^{\text{F}}}{V_{0}}$, that is, it is given by expectation value of the Hamiltonian $\mathfrak{H}$ in the false vacuum state $|0\rangle^{F}$. On the other hand  Eq. (\ref{Fr3})  says that
if $H (t) $ is small then $\frac{\Lambda(t)c^{2}}{3 } = \frac{\Lambda_{0}c^{2}}{3}$ must be even much smaller. Note that until now, no one has proposed   a quantum field theory model predicting a false vacuum state $|0\rangle^{F}$ with the lifetime, $\tau_{0}^{F}$, of the order of the age of the Universe today, $t_{0}$,  (or longer)
such  that the expected value of the operator $\mathfrak{H}$ in this state is correspondingly small.
What is more, there is a high probability in such a case the initial phase of the decay process $[t_{0}^{init}, T_{0})$ can be much longer than a duration of the inflationary phase
with all the consequences of this analyzed in Sec. 3.
So, in the light of this analysis we should rather forget about the possibility that $\tau_{0}^{F} \sim t_{0}$ or $\tau_{0}^{F} > t_{0}$.

\section{Final remarks}

An analysis and estimations presented in the previous Section show that  a universe born in the metastable false vacuum state
has a very high chance of surviving until very late times if the lifetime, $\tau_{0}^{F}$,  of the metastable vacuum is much shorter shorter, than the duration of the the inflation process. In such a case, the universe should survive at least the epoch $t \in (T_{0},T_{1})$ and survive at least until $t > T_{1}$.
  Using inequalities  (\ref{s1}), (\ref{s2}), (\ref{s5}) the fate of the universe in the epoch $t \in (T_{1}, T_{2})$ is difficult to determine because at times
 $t \in  (T_{1}, T_{2})$ the decay rate ${\it\Gamma}(t)$ and $\Lambda(t)$, and hence the parameter $H(t)$, fluctuate and the amplitude of these fluctuations, as well as the time periods during which they reach their local maxima and minima
 depend on the potential $ V(\Phi)$. It is  because the Hamiltonian $\mathfrak{H}$ depends on  $ V(\Phi)$ and  $\mathfrak{H}$ determines the form and properties of the energy density distribution $\omega(E)$ in the false vacuum state $|0\rangle^{F}$.
 In turn knowing $\omega(E)$ and using the equation   (\ref{a-spec}) we can
 find   the survival amplitude ${\cal A}(t)$ and then ${\it\Gamma}(t)$, $E(t)$ and thus  $\Lambda (t)$ too.
Within this scenario, if the universe survived the epoch $t \in (T_{1}, T_{2})$ and  still exists in the epoch $t> T_{2}$, the picture is simpler. In this case the decay rate of the false vacuum state has the form given by Eqs (\ref{g(t)-as}) and  (\ref{Im-h-as}) which allows to rewrite inequalities (\ref{s1}), (\ref{s2}) in a simple form (\ref{s-T2a}), (\ref{s-T2b}) leading to the conclusion that within the considered scenario the universe born in metastable false vacuum state should be  safe.
All these conclusions are important because they also apply to our Universe, as according to the Standard Model (there is no replacement model today),  the Higgs boson and the top quark masses suggest that our Universe was created in the state of metatable vacuum.
Another important conclusion from our analysis is that cosmological models in which the lifetime, $\tau_{0}^{F}$, of a metastable false vacuum is comparable to the age, $t_{0}$, of the Universe or is much longer meet the conditions (\ref{s2}), (\ref{s5}) (see (\ref{sTu})) but such a models do not explain the discrepancy between the measured value of the vacuum energy density and the theoretical one calculated by the field theory methods. Our analysis shows that if the lifetime of the false vacuum state is comparable or longer than the age of the Universe then we are living at the canonical decay times and in such a case  the vacuum energy density, $\varepsilon_{0}^{F}$,  is given by the expected value of the Hamiltonian $\mathfrak{H}$
 in the false vacuum state $|0\rangle^{F}$. On the other hand, in my opinion, the advantage of models with very short lifetime of metastable false vacuum state is that some of them
  may not only satisfy the conditions
  (\ref{s2}), (\ref{s5}) but also they may explain the mentioned discrepancy (see, eg. Figs (\ref{e2}), (\ref{k5}) and analysis presented in \cite{ku2022}).

In general, the above briefly summarized results and conclusions obtained for times $ t < T_{2}$ are quite general and do not depend on the form  of the energy density distribution  function, $\omega (E)$, and thus on $V(\Phi)$ and $\mathfrak{H}$, except that the values of $T_{1}$ and $T_{2}$ depend on the parameters of the model under consideration, i.e. on $\omega (E)$. Only the results obtained and conclusions drawn for times $t > T_{2}$ are based on the assumption that $\omega(E)$ is given by the Breit--Wigner energy density distribution, $\omega (E) \equiv  \omega_{BW}(E)$.
We should be aware that for other $\omega(E)\neq \omega_{BW}(E)$
the inequality (\ref{s-T2gen}) replaces inequalities
inequalities (\ref{s-T2a}), (\ref{s-T2b})
which allows us to study other scenarios for  the epoch $t > T_{2}$ than the scenario described above.

It should be emphasized that the above--analyzed scenario is a consequence of the assumption that the process of the false vacuum decay is a purely quantum decay process
taking place in the gravitational field and is not disturbed by other non--gravitational mechanisms.
Our analysis shows (see Eq. (\ref{s5}) and discussion following it) that forces of the gravity and  forces generated by the positive false vacuum energy density, $\varepsilon_{0}^{F}$, that is  $\Lambda_{0} > 0$, (see eg., Eq. (56) in \cite{ku2022} or \cite{Cheng,Sahni}),  can block the decay of the metastable false vacuum state.
A similar conclusion that gravitation can stabilize false vacuum was drawn in  \cite{Col2}. Namely, therein using completely different quasi--classical formalism it was shown
that the transition from a  state (a false vacuum state) with zero energy, ($\varepsilon_{0}^{F} = 0$ in notations used in this paper) to a state  with negative energy (the true vacuum), which corresponds to $\Lambda < 0$, can be stopped by a gravitation. In other words it was shown therein that in such a case the gravitation can stabilize the false vacuum state and quench its decay.
It should be noted here that the assumption, that the energy in the false vacuum state is zero, $\varepsilon_{0}^{F} = 0$, contradicts the conclusions resulting from  the quantum theory of unstable states: The analysis carried out in Sec. 3 and 4  shows that for quantum unstable states $E(0) \simeq E_{\phi} $, (thus $\varepsilon_{0}^{F} \simeq \varepsilon_{0}^{qft} \neq 0$ to a very good approximation --- see, e.g., Eq. (\ref{E+G-0})). So the  mentioned result presented in \cite{Col2} and discussed here can not describe the real quantum process of the false vacuum decay. The value of that result is only that it shows that gravity can  block the decay process.

The inclusion of other mechanisms and thermal effects in the scenario analyzed in this paper may change some of the above--described details of the this scenario  resulting from the assumption that the decay of the false vacuum is a purely quantum--mechanical decay process.
These changes will manifest themselves  in
variations of the value of
the decay rate ${\it\Gamma}(t)$ of the false vacuum state. As a result of these changes, inequalities (\ref{s2}), (\ref{s5}) may cease to occur or be strengthened depending on the considered model with all the consequences of this. Such mechanisms are discussed in the literature --- see e.g. \cite{Mar,Stojk2} and also \cite{Esp2,JK,De-Ch,Bur1,Bur2,Shk,AS} and others.
In all these papers the standard method of calculations of the decay rate for tunneling was used, i. e. the same approximate method as that in \cite{Coleman:1977py,Callan:1977pt,Col2}. As it was pointed out in \cite{jcap-2020} the decay rate, $\it\Gamma$,  of the metastable false vacuum state obtained within such an semiclassical approximation coincide with the decay rate ${\it\Gamma}_{0}$ proper for the canonical decay times (see discussion in  \cite{jcap-2020}). What is more, analysis presented in Sec. 3 and in \cite{jcap-2020} shows that the decay rate so calculated is not able to reflect correctly early time properties of a decaying quantum system, i.e. at time interval $[t_{0}^{init}=0, T_{0})$, where ${\it\Gamma}(t) \simeq cosnt.\times t$.  So, in fact, all results of a model calculations obtained using such a semiclassical approximation and applied to our Universe mean that in such models the current  epoch lies in the area of canonical decay times for the metastable false vacuum state and, in contrast to the predictions of the quantum theory of unstable states, that the decay process  begins from ${\it\Gamma}(0) = {\it\Gamma}_{0} > 0$ instead of  ${\it\Gamma}(0) =0$.
Let us note that
taking into account early time properties of the decay rate is especially important in the case of models with extremely long lifetimes of the false vacuum because in such cases the early time phase, $[t_{0}^{init} = 0, T_{0})$, of the quantum decay process can be longer than a duration of the inflation process (see Figs (\ref{g1}), (\ref{ge2}), (\ref{g3})).
So, it could therefore have happened that in such models the wrong false vacuum decay rate was used in the analysis of the inflation process.

Last comment:
With reference to the above remarks and from the point of view of the scenarios considered in this paper, it seems that the most promising mechanisms are those that enhance  the decay rate so that the lifetime of the false vacuum is shorter than the duration of the inflation process (e.g. those discussed in \cite{JK,De-Ch,Bur1,Bur2,Shk}).

\hfill\\
\noindent
{\bf The author contribution statement:} The author declares that there are no conflicts of interest
regarding the publication of this article  and that all results presented in this article are the author's own results.

\hfill\\
{\bf Conflicts of Interest}: The author declares no conflict of interest.\\
{\bf Data Availability Statement:} This manuscript has no
associated data, or the data will not be deposited. [Author's
comment: This is a theoretical work and analytical calculations are made. Therefore, no data are required].

\end{document}